%% file: paper.tex
\documentclass[fleqn,usenatbib]{rasti}

\usepackage{newtxtext,newtxmath}

\usepackage[T1]{fontenc}

\DeclareRobustCommand{\VAN}[3]{#2}
\let\VANthebibliography\thebibliography
\def\thebibliography{\DeclareRobustCommand{\VAN}[3]{##3}\VANthebibliography}

\usepackage{graphicx}	
\usepackage{amsmath}	
\usepackage{siunitx}
\usepackage{subfig}

\title[Automation with BiSON:NG]{Automation and Modernisation of Legacy BiSON Solar Observatories with the BiSON:NG Platform}

\author[S.~J.~Hale et al.]{
S.~J.~Hale,$^{1}$\thanks{E-mail: s.j.hale@bham.ac.uk}
E.~Murray,$^{1}$
W.~J.~Chaplin,$^{1}$
Y.~P.~Elsworth,$^{1}$
R.~Howe,$^{1}$
P.~L.~{Pall{\'e}},$^{2}$
E.~J.~{Rhodes},~Jr.$^{3}$
\\
$^{1}$School of Physics and Astronomy, University of Birmingham, Edgbaston, Birmingham B15 2TT, UK\\
$^{2}$Instituto de Astrof{\'i}sica de Canarias, and Department of Astrophysics, Universidad de La Laguna, San Crist{\'o}bal de La Laguna, Tenerife, Spain\\
$^{3}$Department of Physics and Astronomy, University of Southern California, Los Angeles, CA 90089, USA
}

\date{Accepted XXX. Received YYY; in original form ZZZ}

\pubyear{\the\year{}}

\begin{document}
\label{firstpage}
\pagerange{\pageref{firstpage}--\pageref{lastpage}}
\maketitle

\input{abstract}


\input{introduction}
\input{izana}
\input{mtwilson}
\input{conclusions}
\input{acknowledgements}
\input{data}
\input{conflicts}



\bibliographystyle{rasti}
\bibliography{references}







\bsp	
\label{lastpage}
\end{document}

%% file: abstract.tex
%
%
%
%


\begin{abstract}
The Birmingham Solar Oscillations Network (BiSON) provides long-term,
high-precision Sun-as-a-star radial-velocity observations for
helioseismology. While most BiSON sites operate autonomously, two
observatories at Observatorio del Teide, Tenerife, and at Mount Wilson
Observatory, USA, have historically required daily on-site
operation. We present the automation of these sites using the BiSON:NG
next-generation observing platform, incorporating compact fibre-fed
instrumentation and a modular, microservice-based control and data
acquisition architecture.  At both locations, new automated optical
feeds and a common software framework have been deployed, enabling
autonomous operation of the observing subsystems, including guiding,
data acquisition, and instrument control, despite markedly different
physical and operational constraints. The upgrades demonstrate that
modern automation can be successfully integrated within existing and
historic observatory infrastructure. Adoption of the BiSON:NG
architecture reduces operational overhead, mitigates risks associated
with hardware and software obsolescence, and supports sustainable
operation of the network over the multi-decade timescales required for
helioseismic studies. The flexibility of the architecture also
provides a pathway for future deployment of additional small-footprint
observatories.
\end{abstract}



%% file: introduction.tex
%
%
%
%


\section{Introduction}

The Birmingham Solar Oscillations Network (BiSON) makes high-precision
radial-velocity measurements of the Sun-as-a-star.  The network has
operated across six sites since~1992~\citep{Chaplin1996,Hale2016}, and
is currently undergoing a programme of planned infrastructure
modernisation~\citep{Hale2019}.  Four of the sites operate as fully
automated observatories, whilst two -- at Observatorio del Teide,
{Iza\~na}, Tenerife, and at Mount Wilson Observatory, Los
Angeles, USA -- require operators for daily observing.  One aim of the
modernisation programme is to upgrade both of these sites to fully
automated operation.

To support long-term sustainability and reduce operational overhead,
BiSON is implementing a transition to a next-generation observing
platform, hereafter referred to as
BiSON:NG~\citep{10.1117/12.2561282}. BiSON:NG combines a compact,
fibre-fed solar instrument with a modular, microservice-based control
and data acquisition architecture, designed to minimise physical
footprint and cost while enabling fully automated operation and
incremental deployment across legacy sites.

In Section~\ref{sec:izana} we describe the automation of the BiSON
site at Observatorio del Teide, where a new fibre-fed instrument,
automated solar telescope, and weather enclosure have been deployed at
the Solar Pyramid as part of the BiSON:NG upgrade.  In
Section~\ref{sec:mtwilson} we present corresponding work completed at
Mount Wilson Observatory, including installation of a new automated
optical feed within the historic 60-ft Solar Tower and migration of
the site to the BiSON:NG control architecture. Together, these
upgrades demonstrate the deployment of a common, modular automation
framework across legacy observatories operating under markedly
different physical and operational constraints.


%% file: izana.tex
%
%
%
%


\section{Observatorio del Teide, {Iza\~na}, Tenerife}
\label{sec:izana}

\begin{figure*}
  \centering
  \includegraphics[width=\textwidth]{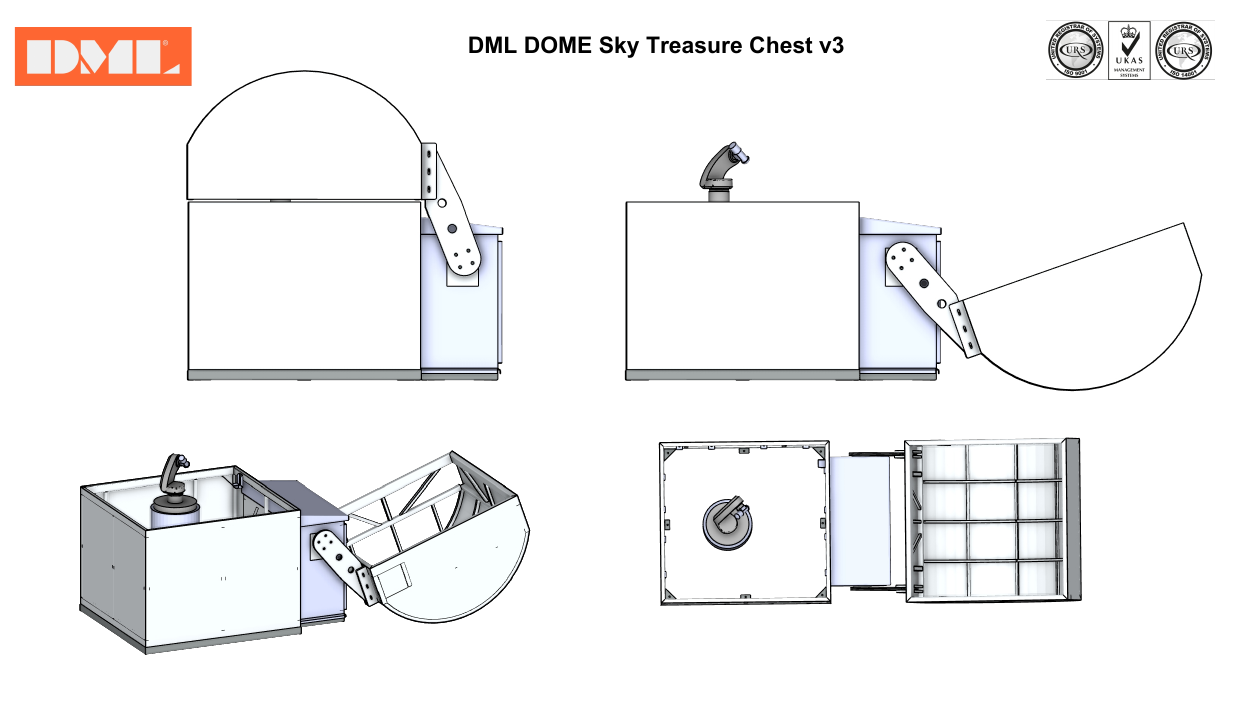}
  \caption{Robotic enclosure –- DML DOME Sky Treasure Chest v3.  A
    free-horizon enclosure designed and manufactured by
    DML~Soluciones~T{\'e}cnicas. Fully constructed in aluminium with
    integrated thermal insulation. The ``treasure chest'' mechanism
    provides a~\SI{180}{\degree} opening, enabling a fully
    unobstructed horizon.  The system includes position-controlled
    opening and closing, acoustic alert signalling during motion, and
    internal humidity control via an integrated dehumidifier.
    Environmental monitoring is provided through humidity, rain, and
    wind sensors, enabling fully automatic operation based on weather
    conditions. The enclosure supports remote control via API,
    allowing seamless integration into automated observatory control
    systems.}
  \label{fig:dml_design}
\end{figure*}

\begin{figure*}
    \centering
    \subfloat{
      \includegraphics[width=0.45\textwidth]{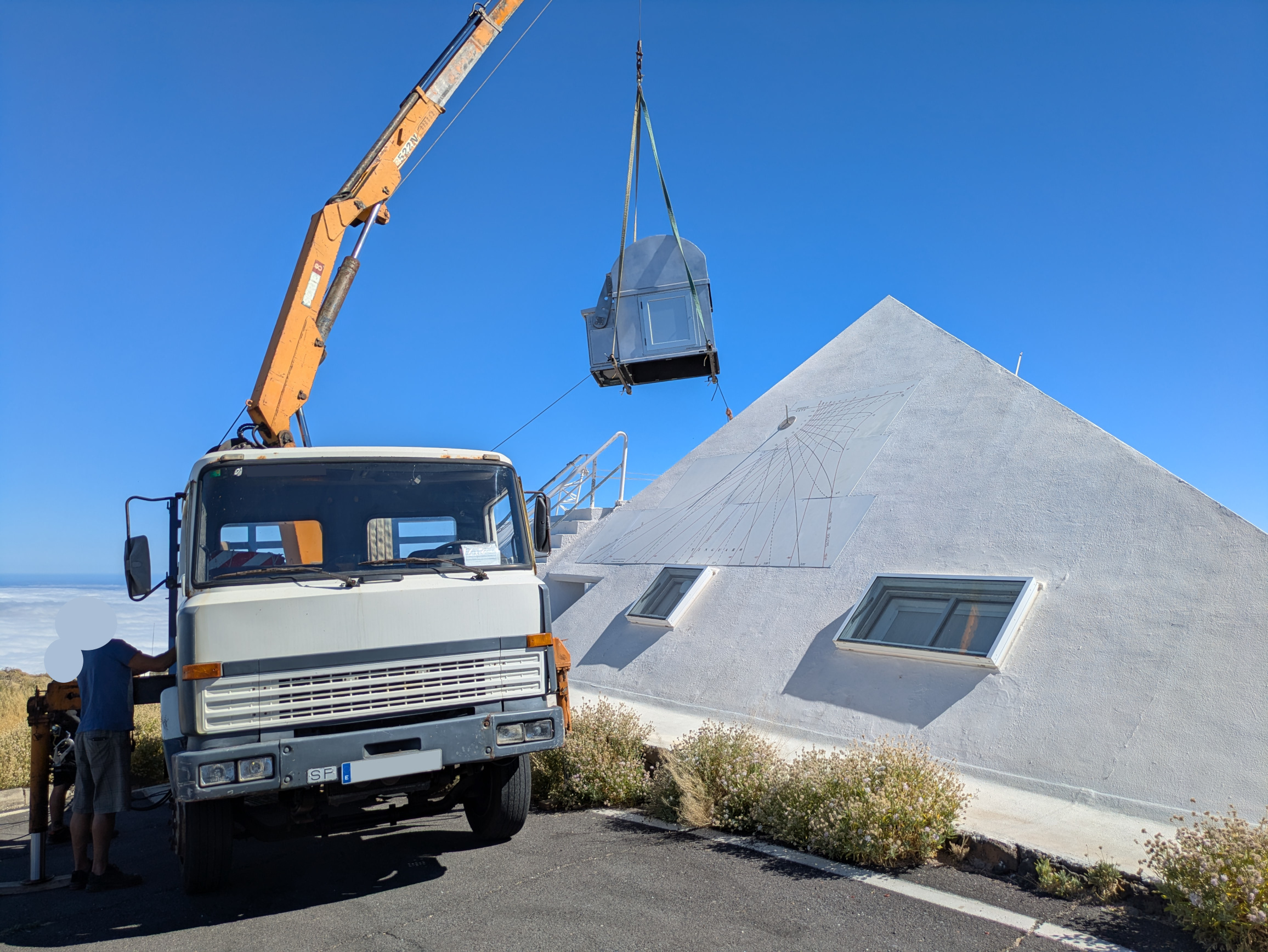}
       \label{subfig:enclosure1}
       }%
    \subfloat{
       \includegraphics[width=0.45\textwidth]{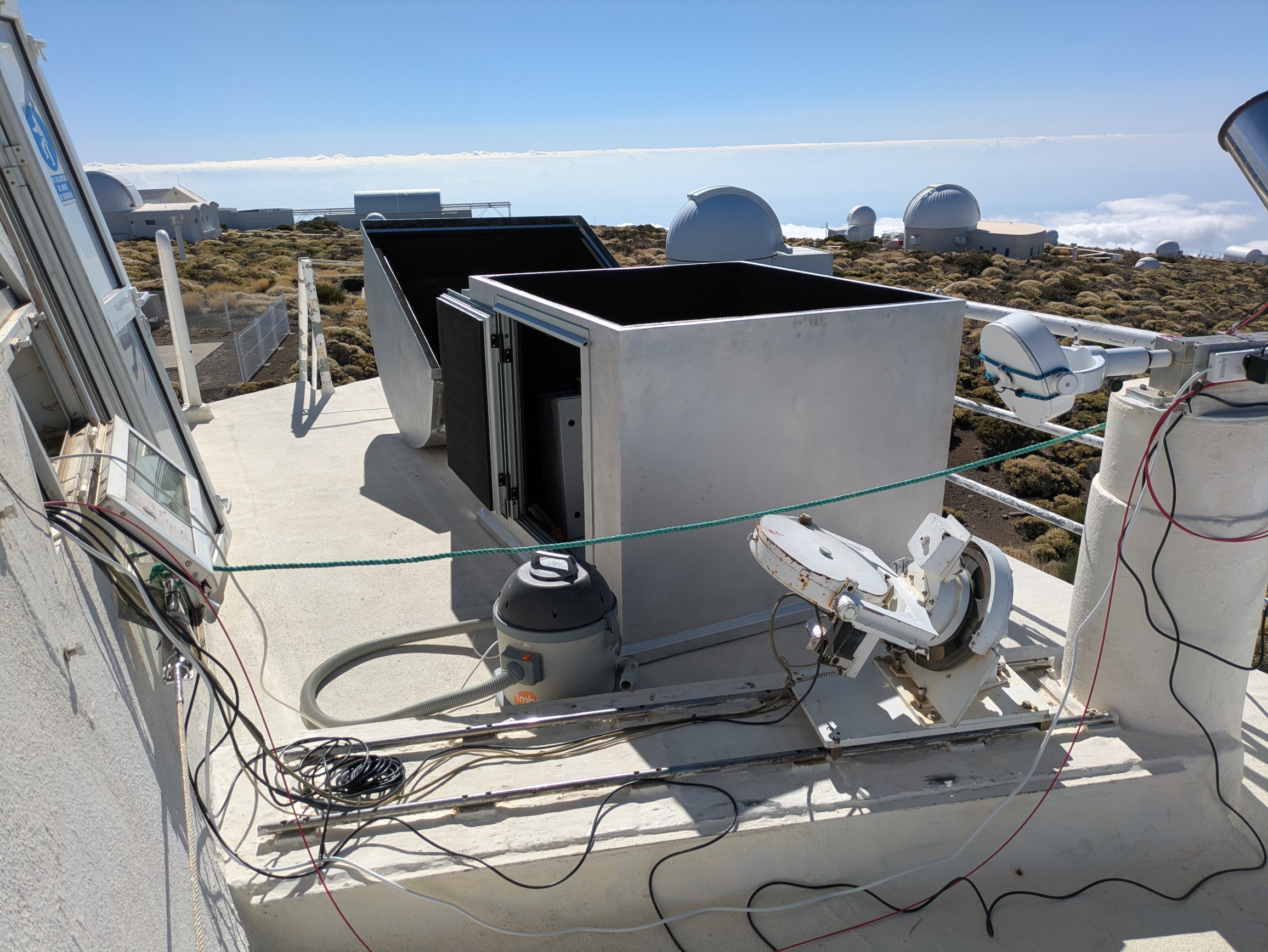}
       \label{subfig:enclosure2}
       }%
    \caption{Left: The enclosure being lifted into place on the upper
      observing platform of the Pyramid.  Right: Installation and
      system commissioning.}
    \label{fig:enclosure}
\end{figure*}

\begin{figure*}
    \centering
    \subfloat{
      \includegraphics[height=0.35\textwidth]{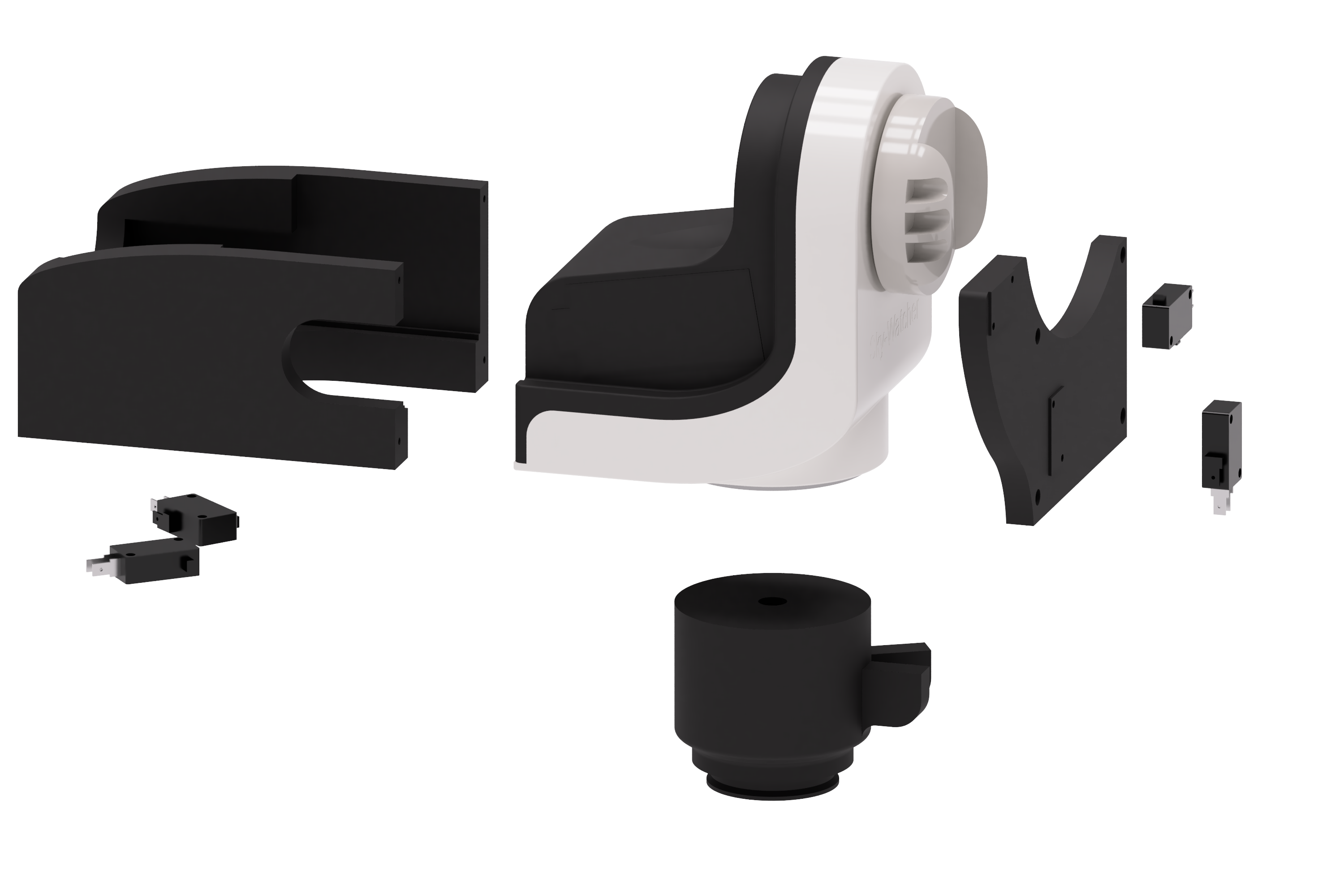}
       \label{subfig:skywatcher1}
       }%
    \vrule width 0.5pt height 0.35\textwidth
    \subfloat{
       \includegraphics[height=0.35\textwidth]{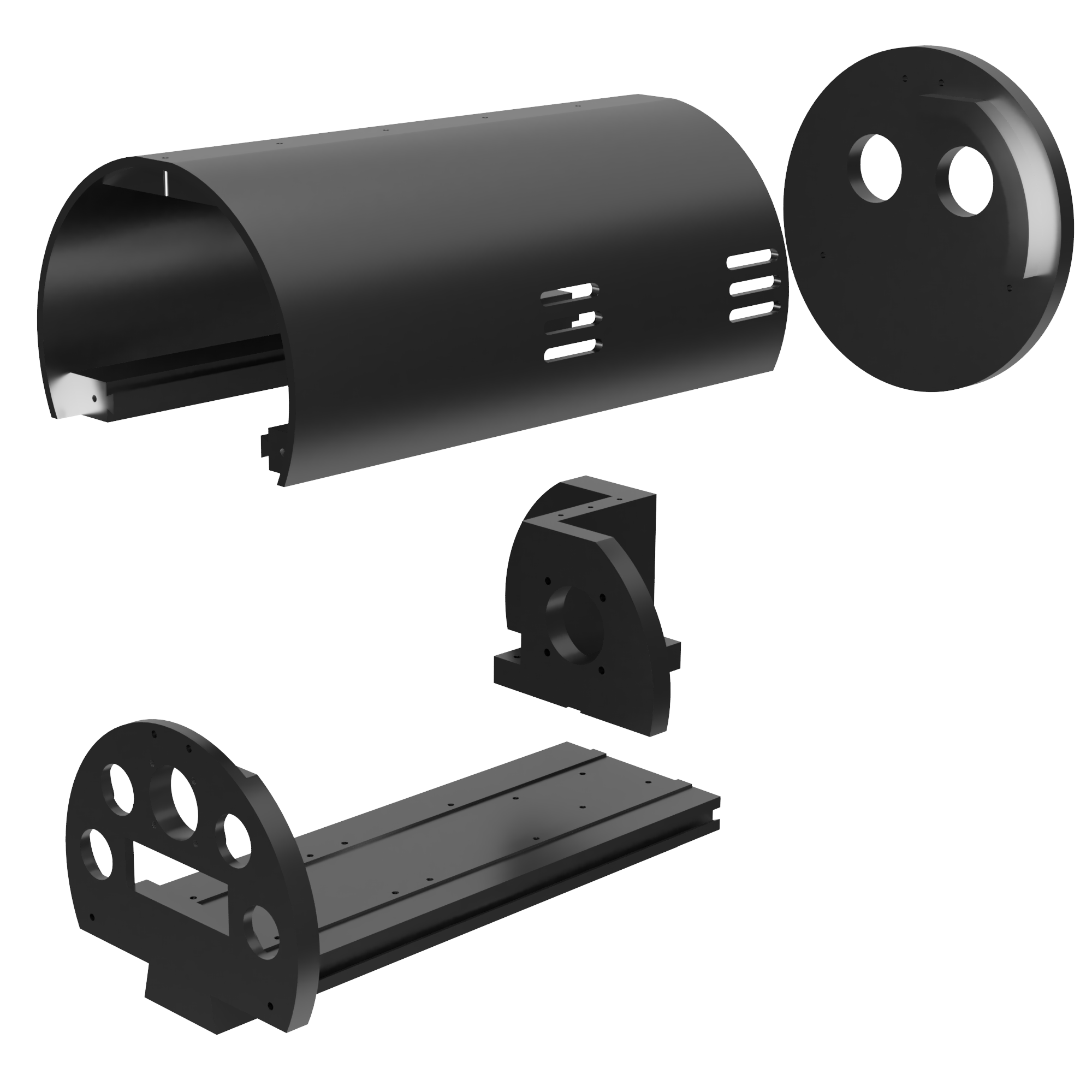}
       \label{subfig:skywatcher2}
       }%
    \caption{Left: An exploded view of a Sky-Watcher AZ~GTi Alt-Az
      mount modified with limit switches for automatic position
      homing.  Right: A custom solar telescope chassis supporting a
      solar fibre-feed, and on-board control and guiding.  Not to
      scale.}
    \label{fig:skywatcher}
\end{figure*}

\begin{figure}
  \centering
  \includegraphics[width=0.4\textwidth]{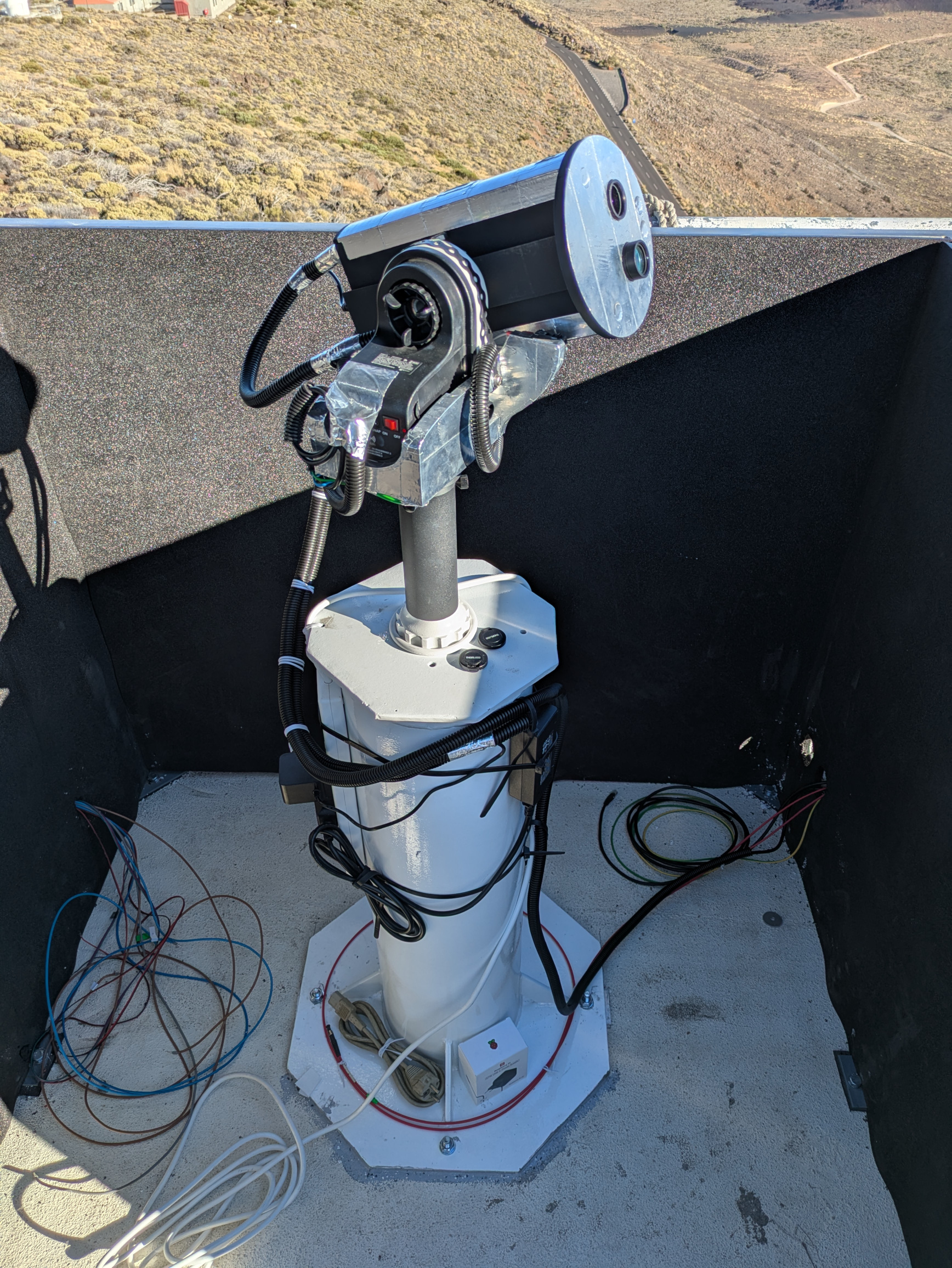}
  \caption{The automated Sky-Watcher AZ~GTi Alt-Az mount and solar
    telescope installed on a pier inside the weather enclosure.}
  \label{fig:telescope}
\end{figure}

\begin{figure*}
  \centering
  \includegraphics[width=0.7\textwidth]{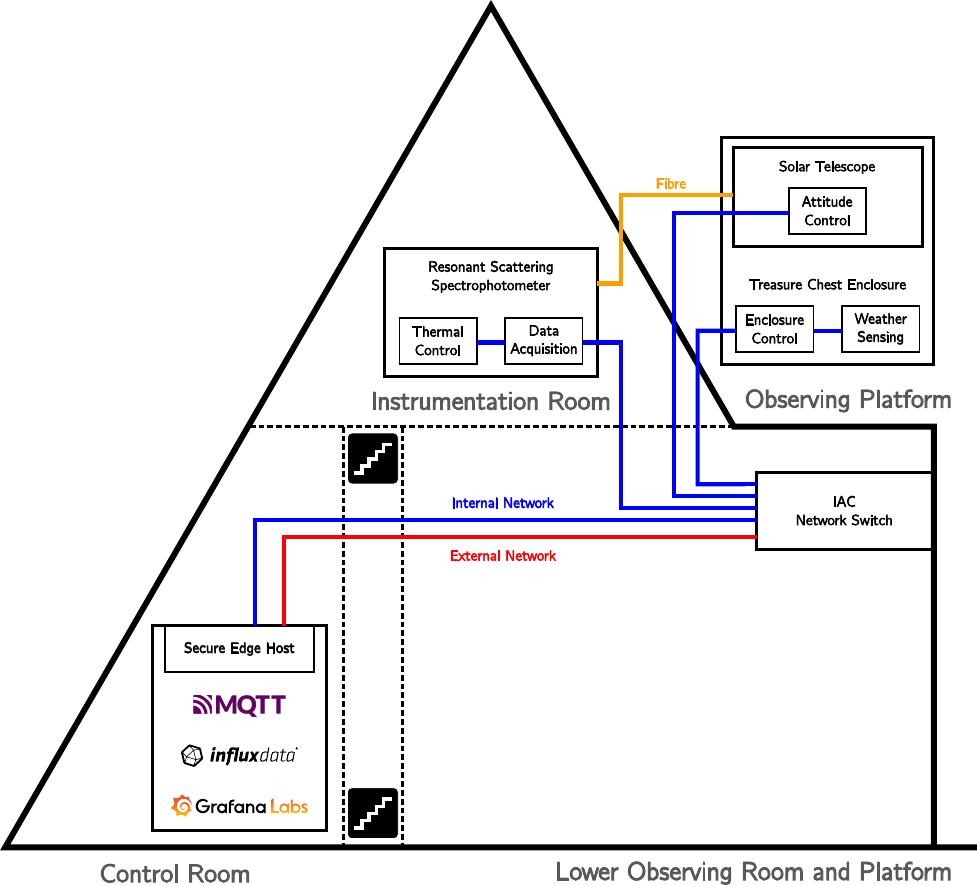}
  \caption{Block diagram of the {Iza\~na} system architecture,
    including approximate physical layout.  The remotely accessible PC
    handling data storage is in the main control room.
    Instrumentation is contained within the upper apex room, with
    fibre-feed from the solar telescope inside the enclosure on the
    external observing platform.  Internal and external network
    connections, indicated by the blue and red lines respectively, are
    separated by VLAN.}
  \label{fig:izana_block_diagram}
\end{figure*}

The Mark-I spectrophotometer at Observatorio del Teide, {Iza\~na},
Tenerife, was first commissioned in~1975 and became the inaugural site
of the Birmingham Solar Oscillations Network in the early
1990s~\citep{Hale2016}.  The instrument operates from the Solar
Pyramid, a tetrahedron-shaped building incorporating two external
observing platforms, each supporting a coelostat. Observations at this
site require an on-site observer to initiate and terminate observing
sessions and to respond to changing weather conditions.

By contrast, the four fully automated BiSON solar observatories employ
a more conventional observatory architecture, consisting
of~\SI{4}{\metre}~domes and large equatorial mounts supporting the
combined telescope and instrument. The upgraded next-generation
network, BiSON:NG, is designed to minimise physical footprint and
cost, simplifying deployment while preserving full automation and
scientific capability~\citep{Hale2019}.

A new ``treasure chest'' weather enclosure, shown in
Figure~\ref{fig:dml_design}, has been installed on the upper observing
platform at the Pyramid, shown in Figure~\ref{fig:enclosure}.  This
enclosure is a fully autonomous system incorporating sensors for rain,
wind, and humidity, and is controlled via a REST API over
Ethernet. The enclosure is commanded to open and close at sunrise and
sunset, but will also close, and re-open, automatically in response to
environmental conditions.  The new architecture incorporates fail-safe
behaviours at multiple levels.  In addition to closing in response to
its onboard environmental sensors independently of the broader
microservice layer, it will also close via watchdog if contact is lost
with the wider system.  On Linux-based devices, service health is
monitored via systemd, which automatically restarts failed
services. At the application level, service liveness is inferred from
message timestamps, with timeouts flagging when expected telemetry is
absent.

The optical feed is based on inexpensive off-the-shelf components,
using a Sky-Watcher~AZ-GTi telescope mount controlled by an on-board
single-board computer and guide camera.  The mount has been modified
with limit switches to enable automatic position homing, and the guide
camera is housed within a custom solar telescope chassis.  Renders of
the model design are shown in Figure~\ref{fig:skywatcher}.  The
combined mount and telescope as installed on the pier inside the
enclosure is shown in Figure~\ref{fig:telescope}.  The new fibre-fed
instrument, currently operating alongside Mark-I in the apex observing
room, is described by~\citet{10.1093/rasti/rzac007}.

The data acquisition and control system has also been
modernised. Legacy systems employ a monolithic architecture in which a
single computer manages all operational functions. In contrast,
BiSON:NG adopts a microservice architecture, in which independent
services are responsible for discrete tasks and encapsulated business
logic -- e.g.,~thermal control, guiding, data acquisition, data
storage, monitoring and alerts, etc.  This allows development,
testing, and updates to be performed on a per-service
basis~\citep{10.1117/12.2561282}.  The modular approach permits
services to run on the most appropriate Ethernet-connected hardware --
whether a PC, microcontroller, or single-board computer -- and reduces
the risk of component obsolescence by allowing individual services to
be replaced without impacting the wider system.

Communication between microservices is handled by MQTT~\citep{mqtt}.
Data are stored using InfluxDB~\citep{influxdb}, and system status is
visualised via a Grafana dashboard~\citep{grafana}.  The control
software is implemented in Python~\citep{10.5555/1593511} on {x86\_64}
platforms, and in {C++}~\citep{cpp} on embedded platforms such as
microcontrollers, reflecting the differing constraints of the target
environments.  A block diagram of the system architecture, together
with a representation of the approximate physical layout, is shown in
Figure~\ref{fig:izana_block_diagram}.  The entrance to the Pyramid is
in the lower observing room.  The PC on the ``edge'' of the network
for remote connection is in the control room.  Upstairs leads to the
instrumentation room housing the spectrophotometer and control
systems, and a further door leads outside onto the upper observing
platform supporting the enclosure and telescope.

The BiSON:NG control architecture was developed to address the
specific requirements of the BiSON instrument suite, including
integration with existing bespoke legacy hardware and support for
embedded and microcontroller platforms alongside conventional
x86~systems. General-purpose observatory control frameworks such as
RTS2~\citep{10.1117/12.672045} were considered; however, the
requirement to support heterogeneous hardware environments including
non-ASCOM-compatible legacy instrumentation, combined with a
preference for minimal external dependencies in a small-team
operational context, led to the adoption of the lightweight
microservice approach built on established open standards (MQTT,
InfluxDB). This approach provides the modularity and replaceability of
components without introducing a large framework dependency on a
codebase outside the team's direct control.

Each BiSON site continues to operate as an independent data
silo. Observing data are stored locally in InfluxDB and retrieved
nightly by an automated process from the University of Birmingham via
SSH. At Birmingham, the data archive is replicated across three
on-site systems, including the University's BlueBEAR HPC facility,
which provides additional storage redundancy. This architecture
ensures that a network outage at any individual site results in at
most one day's data retrieval delay, with no risk of data loss at the
site itself.  As sites are upgraded to the new architecture, real-time
data fusion across sites will become a possibility.


%% file: mtwilson.tex
%
%
%
%


\section{Mount Wilson Observatory, Los Angeles, USA}
\label{sec:mtwilson}

\begin{figure*}
    \centering
    \subfloat{
       \includegraphics[width=0.45\textwidth]{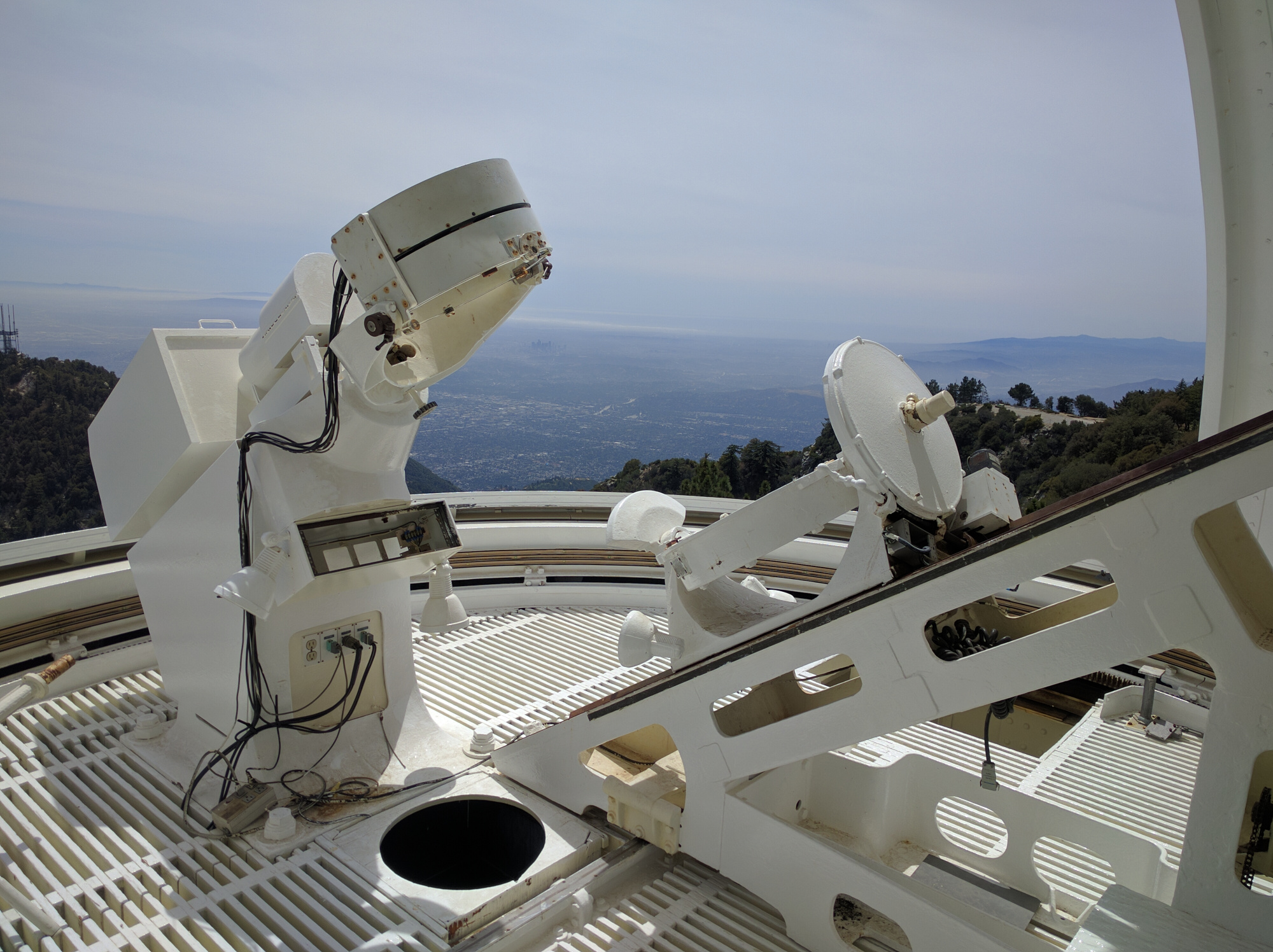}
       \label{subfig:tower1}
       }%
    \subfloat{
       \includegraphics[width=0.45\textwidth]{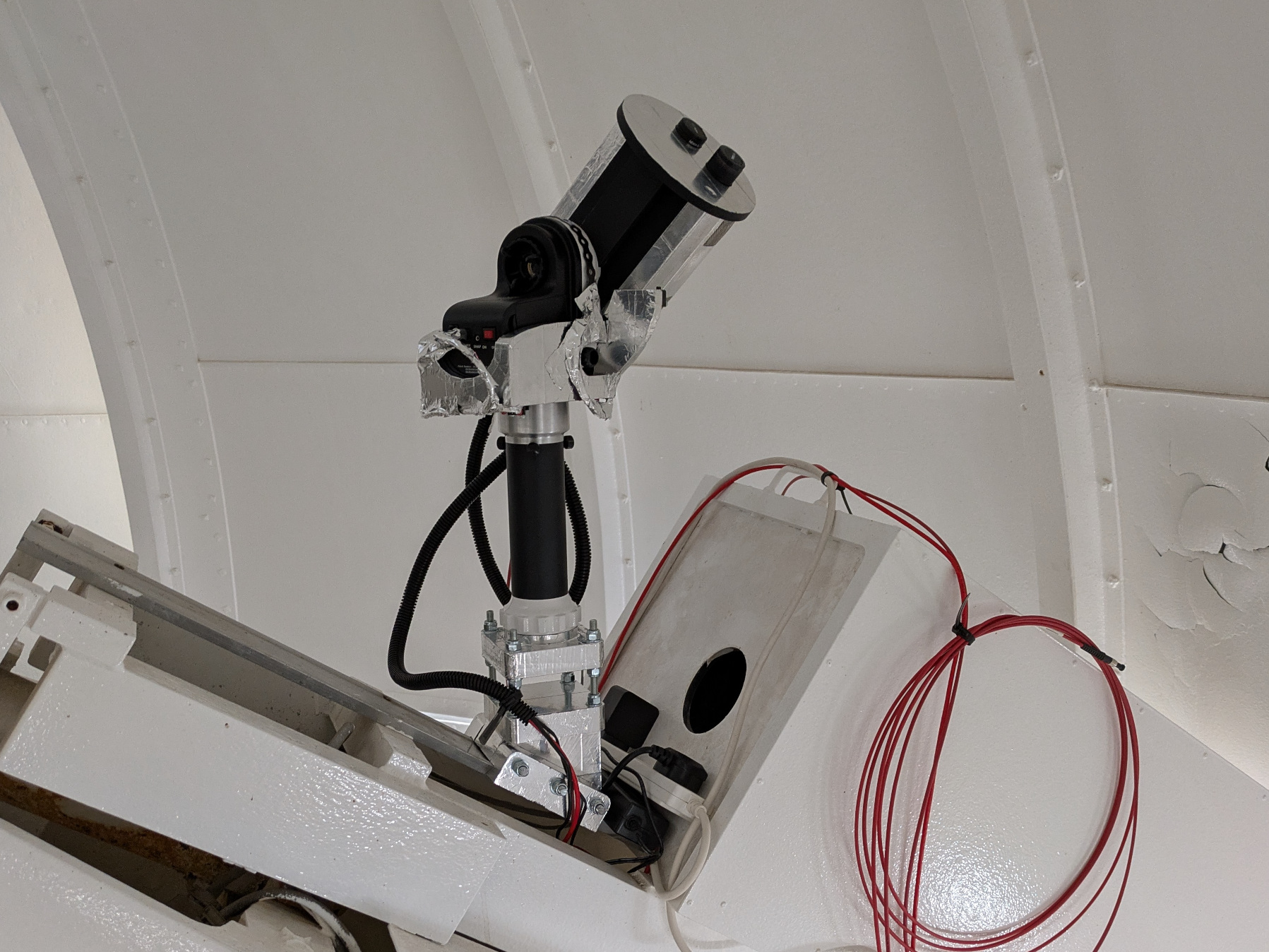}
       \label{subfig:tower2}
       }%
    \caption{Left: The 60-foot Solar Tower Telescope observing
      platform and coelostat.  Right: The automated Sky-Watcher AZ~GTi
      Alt-Az mount and solar telescope installed on the aft-side of
      the second-flat pier.}
    \label{fig:tower}
\end{figure*}

\begin{figure*}
  \centering
  \includegraphics[width=0.7\textwidth]{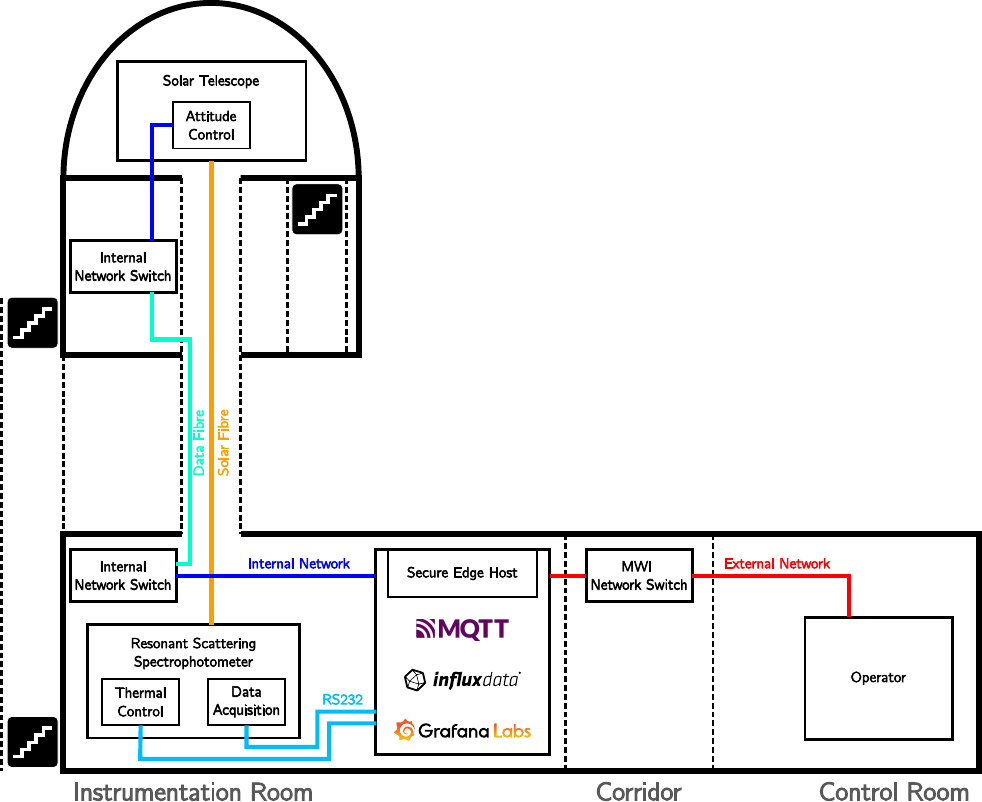}
  \caption{Block diagram of the Mount~Wilson system architecture,
    including approximate physical layout.  The remotely accessible PC
    handling data storage, and the instrumentation, are both contained
    within the main building instrumentation room of the tower.  The
    solar telescope is housed in the observing room at the top of the
    tower, with fibre-feed down the tower shaft to the instrumentation
    room below. Internal and external network connections, indicated
    by the dark blue and red lines respectively, are separated by
    physical network infrastructure.}
  \label{fig:mtwilson_block_diagram}
\end{figure*}

\begin{figure}
  \centering
  \includegraphics[scale=1]{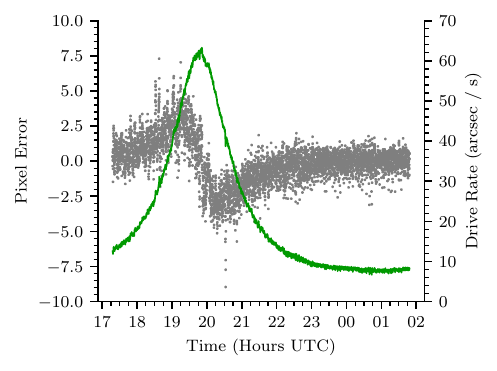}
  \caption{Azimuth axis guiding error.  Pixel drift, indicated by the
    grey dots, is measured on the left axis.  Axis motor drive rate,
    indicated by the green line, is measured on the right axis.  Error
    over the full observing period is~\SI{1.63}{px_{rms}},
    or~\SI{6.6}{\arcsecond_{rms}}.  Over a 3~hour period away from the
    meridian this reduces to~\SI{0.7}{px_{rms}}.  The maximum
    variation over a full day is about~\SI{\pm5}{pixels}.}
  \label{fig:guider_az}
\end{figure}

\begin{figure}
  \centering
  \includegraphics[scale=1]{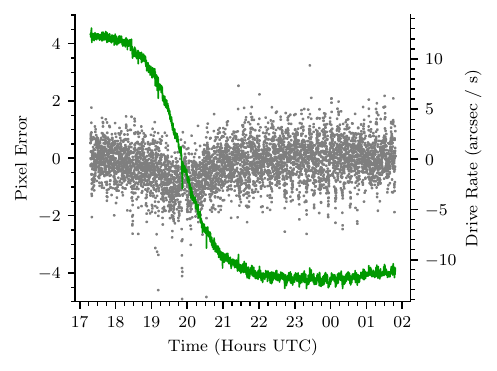}
  \caption{Altitude axis guiding error.  Pixel drift, indicated by the
    grey dots, is measured on the left axis.  Axis motor drive rate,
    indicated by the green line, is measured on the right axis.  Error
    over the full observing period is~\SI{0.74}{px_{rms}},
    or~\SI{3}{\arcsecond_{rms}}.  Over a 3~hour period away from the
    meridian this reduces to~\SI{0.66}{px_{rms}}.}
  \label{fig:guider_alt}
\end{figure}

The first BiSON instrument at Mount Wilson Observatory was installed
in~1992 and subsequently upgraded to the current-generation instrument
in~1996.  The instrument operates from the 60-foot Solar Tower
Telescope, constructed in 1908 and originally used to identify
magnetic fields in sunspots.

As at {Iza\~na}, observations at this site have historically required
an on-site observer to initiate and terminate observing sessions and
to respond to changing weather conditions.  Light is collected by a
coelostat at the top of the tower and directed through the tower shaft
down to the observing room below.  Operation within such historically
significant scientific infrastructure presents a distinct set of
challenges.  In 2018, issues with the coelostat guider electronics
necessitated replacement of the optical feed with a semi-automated
Sky-Watcher SolarQuest solar-tracking mount, together with a long
optical fibre routed down the tower shaft to the observing room. The
instrument was converted to fibre-optic operation using techniques
developed for BiSON:NG~\citep{10.1093/rasti/rzac007}.

The site has now been upgraded to the BiSON:NG standard, mirroring the
configuration at {Iza\~na}, with a fully automated optical feed
installed on the aft-side of the second-flat pier, shown in
Figure~\ref{fig:tower}.  This location provides a wide field of view,
minimising the required dome motion throughout the day, and avoids
shadowing the coelostat, enabling volunteer work to begin on restoring
the original spectroheliostat.  The data acquisition system was also
modified to ensure compatibility with the microservices architecture,
eliminating the need for several legacy digital interface systems.

A block diagram of the system architecture, together with a
representation of the approximate physical layout, is shown in
Figure~\ref{fig:mtwilson_block_diagram}.  The entrance to the Solar
Tower Telescope leads directly into the control room, on the right of
the diagram.  Through a corridor leads to the instrumentation room,
which houses the spectrophotometer and control systems including the
edge PC.  Two new network switches were installed to provide network
connectivity at the top of the tower, required for control of the
guider. This connection is provided via optical fibre installed in the
tower shaft, which is preferred for electrical isolation and
protection against lightning strikes.  The tower observing platform is
accessed via an external staircase, leading to a mezzanine room and
then finally the solar telescope on the upper level.

Figures~\ref{fig:guider_az} and~\ref{fig:guider_alt} show the logged
azimuth and altitude control performance of the guider, respectively,
captured at Mount~Wilson during commissioning on 2026~May~24.  The
guider camera uses a~\SI{6.287}{\milli\metre}
by~\SI{4.712}{\milli\metre} Sony~IMX477 sensor,
with~\SI{1.55}{\micro\metre} square pixels in a~\num{4056}
by~\num{3040}~array.  When coupled with an~\SI{80}{\milli\metre} focal
length objective lens this produces an approximate field of view
of~\SI{4.5}{\degree} by~\SI{3.37}{\degree}, where each pixel has
about~\SI{4}{\arcsecond} field of view.  The Sun has an extent of
about~\SI{32}{\arcminute} and so produces an image about 480~pixels in
diameter on the camera.

The azimuth axis achieves a pixel error of~\SI{1.63}{px_{rms}} over
the full observing period (\SI{6.6}{\arcsecond_{rms}}), or
just~\SI{0.7}{px_{rms}} over a 3-hour period away from the meridian.
The maximum variation over a full day is about~\SI{\pm5}{pixels}.  The
altitude axis performs better due to the smaller dynamic range of
angular velocity.  It is expected this could be improved further with
optimisation of the PID control.

The solar feed uses a~\SI{30}{\milli\metre} focal length objective
lens to couple an image of the Sun into a~\SI{1}{\milli\metre} core
diameter fibre.  The solar image is about~\SI{0.28}{\milli\metre} in
diameter, and the guider error corresponds to a drift of
around~\SI{5.8}{\micro\metre}, comfortably holding the image on the
centre of the fibre.  Performance will be monitored over the following
months to determine how well such consumer-grade mounts respond to
full-time use.

All BiSON observing subsystems at Mount Wilson are now fully
automated, including optical feed, guiding, data acquisition, and
instrument control.  Automation of the tower enclosure, and its
associated weather response, is an active area of development in
collaboration with the Mount Wilson Institute; however, the complexity
of integrating modern automation with the historic tower structure has
proven more challenging than initially anticipated.

The principal challenge is the existing tower enclosure drive system,
which consists of a DC~motor with no positional feedback, controlled
only by open, close, and stop signals. The enclosure comprises two
leaves, one motor-driven and one free-running, the latter being pushed
by the driven leaf on contact, making precise positional control
difficult without hardware modification. Replacing the drive motor
with a digital servo system would provide the positional feedback
required for full automation, but the complexity of this work has
proven prohibitive in the near term. A pragmatic simpler solution is
under development, using optical sensors to detect three fixed
positions corresponding to the required daily movements of the
enclosure. This will provide sufficient positional awareness for
day-to-day automation.  During this period, Mount Wilson Institute
staff manually open and close the tower each day. Once the tower is
opened, routine observing proceeds autonomously without further human
intervention until weather events or sunset.


%% file: conclusions.tex
%
%
%
%


\section{Conclusions}
\label{sec:conclusions}

We have presented the successful automation of the remaining manually
operated BiSON observing sites at Observatorio del Teide, and at Mount
Wilson Observatory. At both locations, new fibre-fed optical systems
and a common BiSON:NG control architecture have been deployed,
enabling autonomous operation of the observing subsystems, including
guiding, data acquisition, and instrument control.

These upgrades demonstrate that full automation of long-running solar
observing facilities can be achieved using inexpensive, commercially
available hardware combined with a modular, microservice-based
software architecture. Crucially, this approach decouples observing
functionality from site-specific infrastructure and hardware
lifecycles, allowing legacy observatories to be modernised
incrementally without wholesale replacement of existing facilities.

Prior to automation, observations at {Iza\~na} and Mount~Wilson
required on-site operators, limiting observing to periods when staff
were available. At Mount~Wilson, operations were typically suspended
over the winter months, with operator availability constraining
coverage during the remaining period. Quantitative assessment of duty
cycle gains relative to the pre-automation baseline will be possible
following an extended period of autonomous operation, and will be
reported in future work.

Adoption of the BiSON:NG architecture substantially future-proofs the
BiSON network against component obsolescence and evolving software
dependencies, supporting sustainable operation over the multi-decade
timescales required for helioseismic studies. This modernisation also
reduces operational overhead and risk, strengthening the case for
continued long-term network operations.

Elements of the BiSON:NG software architecture have additionally been
deployed at the Narrabri site in New South Wales, Australia,
demonstrating that the system can be rolled out independently of
specific optical or mechanical upgrades. This flexibility allows sites
with differing constraints and upgrade pathways to benefit from a
common control and data acquisition framework.

Together, these developments complete a major stage in the
modernisation of the BiSON network and provide a transferable template
for the automation and long-term sustainability of similar small-scale
observatories.  In addition to modernising existing sites, the
BiSON:NG architecture positions the BiSON network to support the
future deployment of small-footprint helioseismic observatories. The
combination of compact instrumentation and a scalable, modular control
framework reduces the technical and operational barriers associated
with establishing new observing locations, without reliance on
site-specific infrastructure.


%% file: acknowledgements.tex
%
%
%
%


\section*{Acknowledgements}

We would like to thank all those who have been associated with BiSON
over the years.  We particularly acknowledge the invaluable technical
assistance at all our remote network sites.

During testing and commissioning at {Iza\~na}: Prof.\,Teodoro Roca,
Antonio Pimienta, all the staff at the Instituto de Astrof\'{i}sica de
Canarias who have contributed to running the {Mark-I} instrument over
many years, and the assistance of the Technical Maintenance team at
the Observatorio del Teide.  DML~Soluciones~T{\'e}cnicas, Canary
Islands, Spain, designed, manufactured, and installed the free-horizon
robotic enclosure and developed its control system.

During testing and commissioning at Mount~Wilson: All former and
current members of the team of {USC} undergraduate observing
assistants, and the assistance of the Mount~Wilson~Institute at
Mount~Wilson~Observatory.

S.J.H, E.M, W.J.C., Y.P.E, and R.H. acknowledge the support of the
United Kingdom Science and Technology Facilities Council (STFC)
through grant ST/V000500/1.

P.L.P acknowledges acknowledges the support (present) of the Spanish
Ministry of Science, Innovation and and Universities under the grant
PID2023-146453NB-100.

E.J.R. acknowledges financial support provided by the BiSON Project to
the University of Southern California (USC) for routine maintenance
and repair to the BiSON data acquisition hardware. E.J.R. also wishes
to acknowledge financial support provided over the past 23~years for
undergraduate observing assistants by the following USC student
support programs: the Provost's Undergraduate Research Fellowship
(PURF), the Undergraduate Research Associates Program (URAP), the USC
Dornsife Student Opportunities for Academic Research (SOAR) Program,
the USC Dornsife Summer Undergraduate Research Fund (SURF) Program,
and the Lick Scholarship Fund of the USC Department of Physics and
Astronomy. E.J.R. also wishes to acknowledge the Mount Wilson
Institute for providing on-going access to the 60-Foot Solar Tower.



%% file: data.tex
%
%
%
%


\section*{Data Availability}

All data are freely available from the {BiSON} Open Data
Portal~--~\url{http://bison.ph.bham.ac.uk/opendata}.  Data products
are in the form of calibrated velocity residuals, concatenated into a
single time series from all {BiSON} sites.  Individual days of raw or
calibrated data, and also bespoke products produced from requested
time periods and sites, are available by contacting the authors.


%% file: conflicts.tex
%
%
%
%


\section*{Conflicts of Interest}

Authors declare no conflict of interest.
